\documentclass[aps,pre,twocolumn]{revtex4}

\usepackage{graphicx,amssymb,amsmath}
\usepackage{color} 
\usepackage{multirow}

\newcommand{\be}{\begin{equation}}
\newcommand{\ee}{\end{equation}}

\newcommand{\bea}{\begin{eqnarray}}
\newcommand{\eea}{\end{eqnarray}}

\def\eq#1{Eq.~(\ref{#1})}
\def\eqs#1#2{Eqs.~(\ref{#1},\ref{#2})}
\def\fig#1{Fig.~\ref{#1}}

\usepackage{array}
\newcolumntype{C}[1]{>{\centering\let\newline\\\arraybackslash\hspace{0pt}}m{#1}}
\newcolumntype{L}[1]{>{\raggedright\let\newline\\\arraybackslash\hspace{0pt}}m{#1}}
\newcolumntype{R}[1]{>{\raggedleft\let\newline\\\arraybackslash\hspace{0pt}}m{#1}}

\begin{document}

\title{dsDNA persistence length with divalent ions}

\author{Manoel Manghi}
\email{manoel.manghi@univ-tlse3.fr}
\author{Audrey Denis}
\author{Nicolas Destainville}
\affiliation{Laboratoire de Physique Th\'eorique, Universit\'e de Toulouse, Universit\'e de Toulouse III - Paul Sabatier, CNRS, France}
\author{Emmanuel Trizac}
\affiliation{Universit\'e Paris-Saclay, CNRS, LPTMS, Orsay, France}
\author{Catherine Tardin}
\email{catherine.tardin@ipbs.fr}
\affiliation{Institut de Pharmacologie et Biologie Structurale (IPBS), Universit\'e de Toulouse, CNRS, Universit\'e de Toulouse III - Paul Sabatier, Toulouse, France}

\date{\today}

\begin{abstract}
Finding a theoretical formula for the persistence length of polyelectrolytes for the whole experimental range of salt concentration is a long standing challenge. Using the Tethered Particle Motion technique, the double-stranded DNA persistence length is measured for four monovalent and divalent salts on a three-decade concentration range. The formula proposed by Trizac and Shen [\textit{EPL}, \textbf{116} 18007 (2016)] and extended to divalent ions fits the data. This formula mixes the high salt limit solution of the Poisson-Boltzmann equation together with the DNA charge renormalisation. Magnesium ions induce a fitted DNA radius smaller than the geometrical one, consistent with a site-specific binding.
\end{abstract}

\maketitle

The variation of the persistence length of polyelectrolytes $L_p$ as a function of the salt concentration is both an experimental and theoretical issue. Experimentally, $L_p$ has been obtained for high salt concentrations using birefringence experiments since the 70's by measuring the DNA radius of gyration~\cite{Harrington1978}. Then, several force-free techniques have been proposed to increase the resolution without real improvement (see~\cite{Savelyev2012,Macromolecules}). Other types of experiments (especially applied to flexible biopolymers) consist in stretching the polyelectrolyte using magnetic tweezers~\cite{Baumann1997,Saleh2009,Innes-Gold2021}.

Recently, some of us measured the double-stranded DNA (dsDNA) persistence length using the  single-molecule technique of high-throughput tethered particle motion (HT-TPM)~\cite{Macromolecules,PRL2019}. It allowed us to measure a persistence length between 120 and 40~nm for an unprecedentedly large salt concentration range $[10^{-3}, 4]$~mol/L. Both monovalent salts (LiCl, NaCl, and KCl) and salt with divalent counterions (Mg$^{2+}$, Ca$^{2+}$ and Putrescine) were studied experimentally. Although theoretical expressions fitted successfully these data, no unique formula was able to fit both monovalent and divalent salt cases. In this Letter, we propose a unique theoretical framework which allows us to fit new and experimental data using 4 types of salts with the following valences $(z_-,z_+)=(1:1)$, $(1:2)$, $(2:1)$ and $(2:2)$.

The monovalent $(1:1)$ salt (NaCl) case was very well fitted~\cite{PRL2019} by the formula proposed by Trizac and Shen in 2016~\cite{TS}. This formula is to date the most elaborated version of the concept of electrostatic persistence length, first developed by Odijk, Skolnick and Fixman (OSF) in 1977~\cite{Odijk,SF}. It consists in separating the full persistence length in two contributions, the bare polyelectrolyte persistence length obtained at infinite salt concentration, $L_{p0}$, and the electrostaticcontribution, $L_{\rm el}$,
\be
L_p=L_{p0}+L_{\rm el}=L_{p0}+\frac{1+4\kappa a}{4 \ell_{B}\kappa^{2}}[\xi_{\rm eff}(\kappa a,\xi)]^2,
\label{Lel}
\ee
where $a$ is the polyelectrolyte radius. The Bjerrum length $\ell_B=e^2/(4\pi\epsilon_0\epsilon_{\rm w}k_BT)$ is equal to 0.7~nm at room temperature ($T=300$~K, $k_B$ is the Boltzmann constant) in water ($\epsilon_{\rm w}=78$, $e$ is the electron charge, and $\epsilon_0$ the permittivity of the vacuum). The Debye inverse screening length is $\kappa=(8\pi\ell_B I)^{1/2}$ where 
\be
I=\frac12(z_+^2 c_+ + z_-^2 c_-)
\ee 
is the ionic strength ($c_\pm$ are the anion and cation concentrations, and $z_\pm$ their valence). 

\begin{figure}[t]
\begin{center}
\includegraphics[height=5cm]{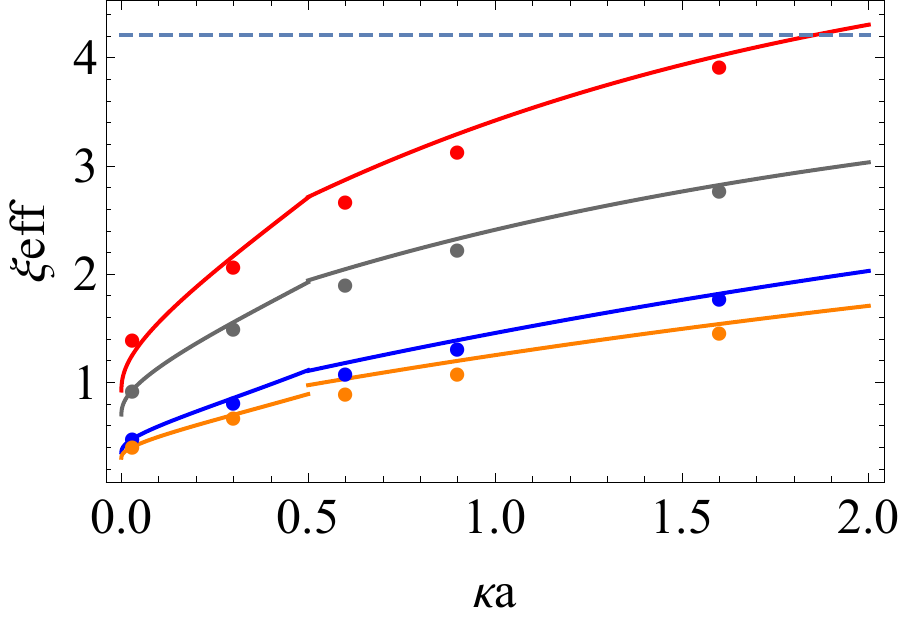}
\includegraphics[height=5cm]{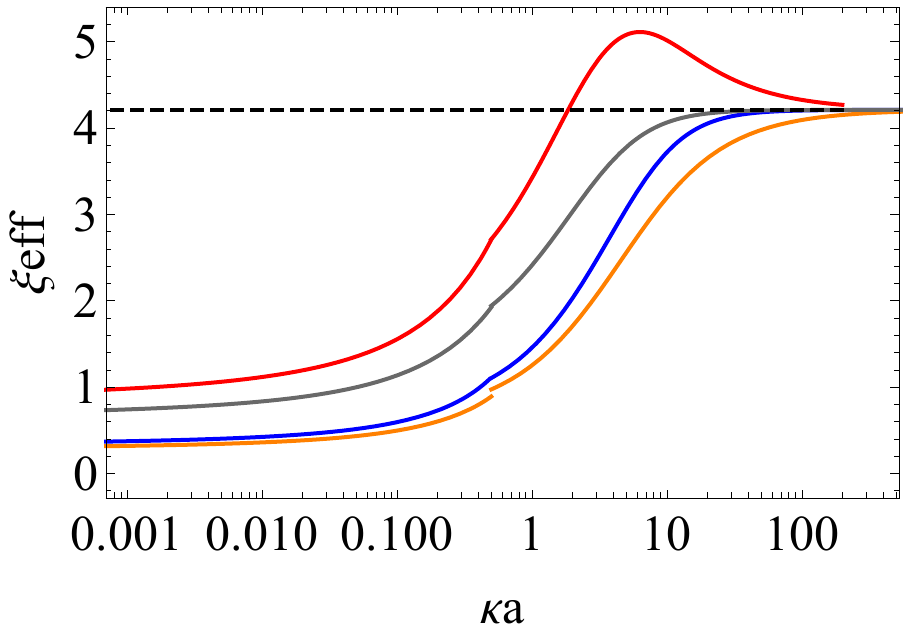}
\caption{Top: Comparison between the analytical expressions of $\xi_{\rm eff}(\kappa a,\xi_{\rm DNA})$ given in \eqs{xieff 1.1>}{xieff 1.1<} (solid lines) and the numerical estimate (dots). Colors corresponds to the different types of ions: red $(2:1)$, grey $(1:1)$, blue $(2:2)$, and orange $(1:2)$. Bottom: Log-linear plot of  $\xi_{\rm eff}(\kappa a,\xi_{\rm DNA})$ for the same parameters on a wider range of $\kappa a$. The dashed horizontal line corresponds to the bare DNA density charge $\xi_{\rm DNA}=4.21$ reached at infinite salt concentration.}
\label{f1}
\end{center}
\end{figure}

The parameter $\xi_{\rm eff}(\kappa a,\xi)$ is the effective dimensionless linear charge density and is function of the ratio between the polyelectrolyte radius $a$ and the Debye length, $\kappa^{-1}$, and the bare dimensionless linear charge density $\xi = \ell_B/A$ where $A$ is the average distance between charges along the polyelectrolyte. It is formally defined as the effective charge felt by the electrostatic potential in the Debye-H\"uckel (DH) approximation at long distances $r\gg\kappa^{-1}$~\cite{Alexander1984}. For monovalent salts, $\xi_{\rm eff}(\kappa a,\xi)\leq\xi_{\rm DNA}=4.21$ for dsDNA, where the equality is obtained at very high screening, $\kappa a\gg1$ (see the grey curve in \fig{f1} bottom). 
Determining the expression of  $\xi_{\rm eff}(\kappa a,\xi)$ is an issue, which has been solved using limiting expressions of the solution of the Poisson-Boltzmann equation~\cite{ka<1/2,ka>1/2}. The fact that $\xi_{\rm eff}$ differs from the bare value $\xi$ is due to the condensation of counterions along the DNA, as first noticed by Manning~\cite{Manning1969} and Oosawa~\cite{Oosawa}.

The well known OSF formula for the persistence length is obtained by replacing $\xi_{\rm eff}(\kappa a,\xi)$ in \eq{Lel} by its limiting value at high salt $\xi_{\rm eff}(\kappa a\to\infty,\xi)\to\xi$ and neglecting the $4\kappa a$ term~\cite{Odijk,SF}. The Manning-Oosawa correction to the OSF formula is obtained by using a constant $\xi_{\rm eff}(\kappa a,\xi)=1/z_+$, independently of $\kappa$~\cite{Manning1969,Oosawa}. It has been shown recently that these two formula do not fit the experimental data on a large range of salt concentration~\cite{Macromolecules}. 

A well defined expression of $\xi_{\rm eff}$ has been obtained using exact asymptotic expansions of the solution of the cylindrical Poisson-Boltzmann equation,
\be 
\Delta\Psi=\frac{\kappa^2}{z_{+}+\vert z_{-}\vert}\left(e^{\vert z_{-}\vert\Psi}-e^{-z_{+}\Psi}\right),
\label{PB}
\ee
where $\Psi=e\phi/(k_BT)$ is the dimensionless electrostatic potential, with the boundary conditions
\be
\left.\frac{\partial\Psi}{\partial r}\right\vert_{a}= \frac{2\xi}{a}, \qquad{\rm and}\qquad \lim_{R\to\infty}\Psi(R)=0.
\label{CB2}
\ee

The solution of \eq{PB} is known to behave, at large distances $\kappa r\gg1$, like the solution of the high salt limit version of \eq{PB}, the linearized DH equation $\Delta\Psi=\kappa^2\Psi$. Hence the electrostatic potential simplifies to
\be
\Psi\simeq-  2\xi_{\rm eff} \frac{K_{0}(\kappa r)}{\kappa aK_{1}(\kappa a)} \qquad{\rm for}\qquad  r\gg\kappa^{-1},
\label{PsiDH}
\ee
where $K_{0}$ and $K_{1}$ are modified Bessel functions of 2nd kind.

For $(1:1)$ electrolytes, the effective charge density is given~\cite{TS} in the low salt limit ($\kappa a<\frac12$) by:
\bea
\xi^{(1:1)}_{\rm eff}(\kappa a,\xi) &=& \frac{2}{\pi}\kappa aK_{1}(\kappa a)\cosh[\pi \mu^{(1:1)}(\kappa a,\xi)],
\label{xieff 1.1<}\\
\mu^{(1:1)}(\kappa a,\xi) &=& \frac{\pi}{2\left[\ln(\kappa a)+\gamma-\ln8-\frac{1}{\xi-1}\right]}.
\label{mu11}
\eea
In the high salt limit ($\kappa a>\frac12$, chosen such as the two functions intersect), we have
\bea
\xi^{(1:1)}_{\rm eff}(\kappa a,\xi)&=&\left[2\kappa a+\frac{1}{2}\left(5-\frac{\left[T\left(\frac{\xi}{\kappa a +1/2}\right)\right]^{4}+3}{\left[T\left(\frac{\xi}{\kappa a +1/2}\right)\right]^{2}+1}\right)\right]\nonumber\\
& & \times T\left(\frac{\xi}{\kappa a +1/2}\right),
\label{xieff 1.1>}
\eea
where $T(x)=(\sqrt{x^{2}+1}-1)/x$. On checks that $\xi_{\rm eff}(\kappa a\to\infty,\xi)\to\xi$.
The $(1:1)$ case has been very well fitted by \eq{Lel} for DNA in monovalent NaCl salt in~\cite{PRL2019} with the fitting parameter values $L_{p0}=41$~nm and $a=0.85$~nm.

The expression $\xi_{\rm eff}(\kappa a,\xi)$ for the case of divalents co-ions and divalent counter-ions are slightly different from \eqs{xieff 1.1>}{mu11} and have been computed in Refs.~\cite{ka<1/2,ka>1/2}.
In the case of a $(1:2)$ salt and $\kappa a<\frac12$,
\bea
\xi_{\rm eff}^{(1:2)}&=& \frac{\sqrt{3}}{2\pi}\kappa aK_{1}(\kappa a)[2\cosh(\pi \mu^{(1:2)}(\kappa a, \xi))-1],
\label{xieff 1.2<}\\
\mu^{(1:2)}&=& \frac{\pi}{3\left[\ln(\kappa a)+\gamma-\frac13\ln2 -\frac32\ln3-\frac{1}{2\xi-1}\right]},
\eea
and for $\kappa a>\frac12$, it is obtained from a multiple scale expansion~\cite{ka>1/2}
\be
\xi_{\rm eff}^{(1:2)}(\kappa a, \xi)= A\left[3\kappa a+\frac98-\frac{c_{1}}{2}+{\cal O}\left(\frac{1}{\kappa a}\right)\right],
\label{xieff 1.2>}
\ee 
with $s={2\xi}/{(\kappa a)}$ and
\bea
A &=& \frac{1}{s}\left[-2+s+2^{\frac{3}{2}}\left(2-s+s^{2}\right)^{\frac{1}{2}}\cos\left(\frac{\theta+4\pi}{3}\right)\right], \label{A}\\
\theta &=& \cos^{-1}\left(\frac{-4+3s-3s^{2}+s^{3}}{\sqrt{2}(2-s+s^{2})^{\frac{3}{2}}}\right), \label{theta}\\
c_{1} &=& \frac{2A^{6}-12A^{5}+37A^{4}+170A^{3}-24A^{2}-74A+9}{4(1-2A+6A^{2}-2A^{3}+A^{4})}.\nonumber\\\label{c1}
\eea
In the case of a $(2:1)$, we have for $\kappa a<\frac12$
\bea
\xi_{\rm eff}^{(2:1)}&=&\frac{\sqrt{3}}{2\pi}\kappa aK_{1}(\kappa a)[2\cosh(\pi \mu^{(2:1)}(\kappa a, \xi))+1],
\label{xieff 2.1<}\\
\mu^{(2:1)} &=&\frac{2\pi}{3\left[\ln(\kappa a)+\gamma-\ln2-\frac32\ln3-\frac{1}{\xi-1}\right]},
\eea
and for $\kappa a>\frac12$, it is given by \eq{xieff 1.2>}--\eq{c1} by changing $s\to-s$ and $\theta+4\pi/3\to\theta$.
Note that, at high salt $\kappa a\gg1$, $\xi_{\rm eff}(\kappa a,\xi)\simeq \xi\pm\frac23 \frac{\xi^2}{\kappa a}$ where $+$ is for $(2:1)$ and $-$ for $(1:2)$. In particular one has $\xi_{\rm eff}>\xi$ for the $(2:1)$ electrolyte, i.e. when co-ions are divalent.

For the $(2:2)$ electrolyte, we have deduced the formula for $\xi_{\rm eff}$ from the one for the $(1:1)$ case as follows. If we set $\tilde \Psi= 2\Psi$, one obtains for the $(2:2)$ case the same Poisson-Boltzmann equation as for the $(1:1)$ case. The only difference comes from the boundary condition (Gauss law) $\left.\frac{\partial \tilde \Psi}{\partial r}\right\vert_{a}=\frac{4\xi}{a}$, i.e. with a doubled fictive charge density. Hence to obtain $\xi_{\rm eff}$ in the  $(2:2)$ case, we just have to replace $\xi$ by $2\xi$ in \eqs{xieff 1.1>}{mu11} and to divide $\xi_{\rm eff}$ by 2:
\be
\xi^{(2:2)}_{\rm eff}(\kappa a,\xi)=\frac12\xi^{(1:1)}_{\rm eff}(\kappa a,2\xi).
\ee
Here again, the non-linearity of $\xi^{(1:1)}_{\rm eff}(\kappa a,\xi)$ with $\xi$ leads to $\xi^{(1:1)}_{\rm eff}>\xi^{(2:2)}_{\rm eff}$.

First we have checked numerically the validity of these formula for the effective charge density $\xi_{\rm eff}(\kappa a, \xi)$. To do so, we have solved numerically \eq{PB} using the shooting method~\cite{NR}, then we have adjusted $\xi_{\rm eff}(\kappa a, \xi_{\rm DNA})$ defined in \eq{PsiDH} by superimposing the DH solution to the exact numerical solution for large $r$. The numerical values are reported in \fig{f1} top as dots for various salt conditions: $(1:1)$ in gray, $(2:1)$ in red, $(1:2)$ in orange, and $(2:2)$ in blue. The agreement with the analytical asymptotic formula (in solid lines) is very good. For intermediate $\kappa a$ values, the numerical values are slightly lower than the analytical ones (with a maximal difference of 8\%). One observes the cusp at $\kappa a=1/2$ which corresponds to the junction between the two formulas \eqs{xieff 1.1>}{xieff 1.1<}. For $\kappa a>2$, the agreement between \eq{xieff 1.1>} and exact numerical values is better and better as checked in~\cite{ka>1/2}. The function $\xi_{\rm eff}(\kappa a, \xi_{\rm DNA})$ is plotted in log-linear in \fig{f1} bottom for $10^{-3}\leq \kappa a\leq 10^3$. It is interesting to note that although this function is monotonous for the $(1:1),(1:2),(2:2)$ cases, it shows a maximum slightly larger than $\xi_{\rm DNA}$ around $\kappa a\simeq 6$ for the $(2:1)$ case. This overshooting effect has already been observed in Ref.~\cite{ka>1/2} and occurs because of non-linear effects: the number of coions that condense along the DNA is larger than half of the monovalent counterions thus inducing $\xi_{\rm eff}>\xi_{\rm DNA}$. This does not occur for the $(1:2)$ case because the entropy cost (on the order of $k_BT$ per ion) to screen the DNA charge by divalent counterions remains the same whereas the electrostatic energetic gain is doubled.
Note that the experimental range of ionic strengths studied in this work corresponds to $0.074\leq\kappa a\leq2.33$ which is the intermediate range where $\xi_{\rm eff}(\kappa a, \xi_{\rm DNA})$ strongly increases, far from the asymptotic values reached for very low and very high $\kappa a$. Moreover, the bare charge density $\xi$ is reached for $\kappa a> 20$ for NaCl, which is far beyond the accessible experimental values. For instance, the solubility limit is $I\simeq7$~mol/L for NaCl corresponding to $\kappa a\simeq 9$.
\begin{figure}[t!]
\begin{center}
\includegraphics[width=0.9\columnwidth]{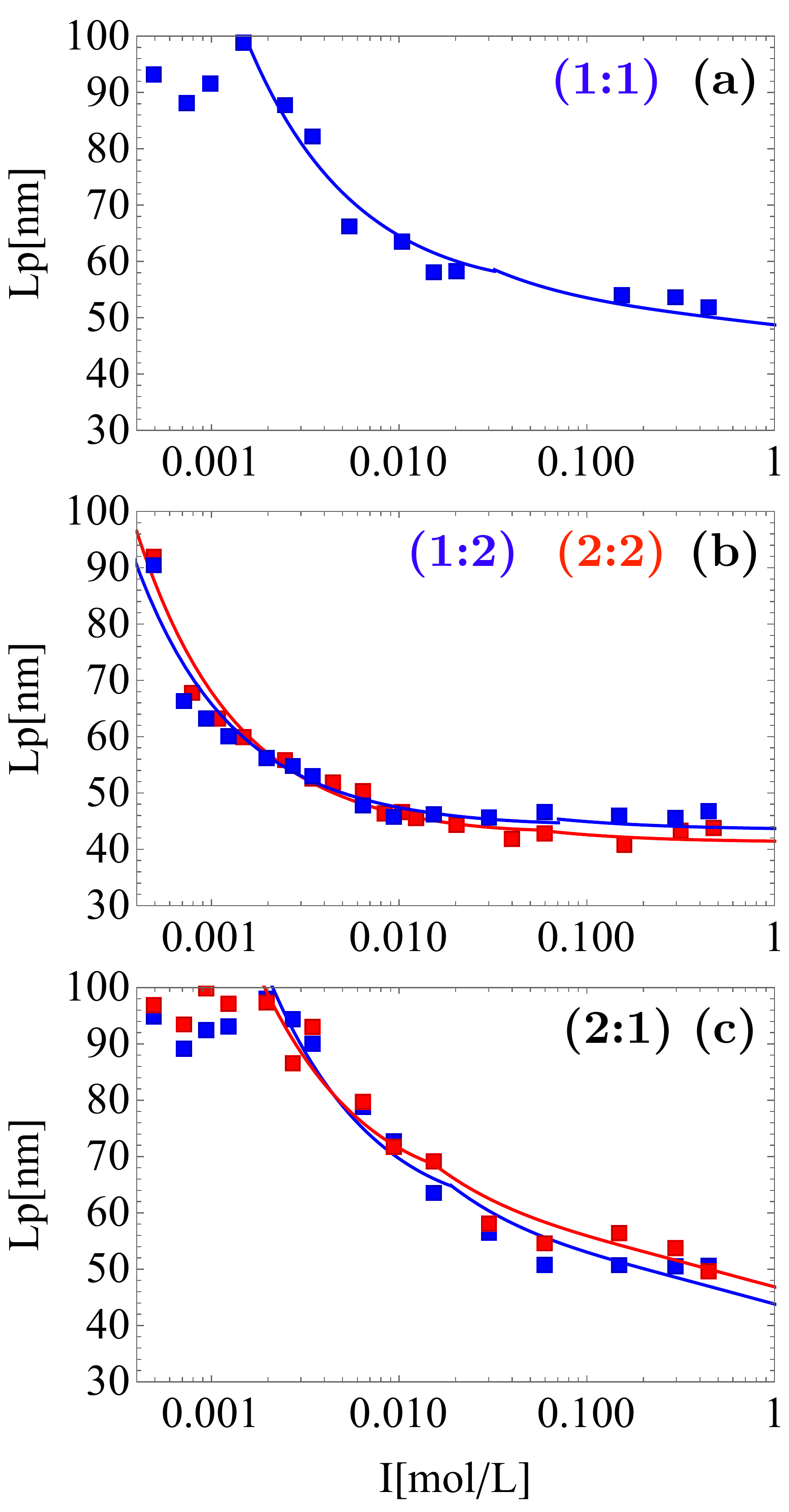}
\caption{dsDNA persistence length $L_p$ vs. ionic strength $I$. Squares are experimental data and solid lines are fitted theoretical expressions such as \eqs{xieff 1.1>}{xieff 1.1<}. \textbf{(a)} $(1:1)$ NaCl salt. The fitting parameters are the bare persistence length $L_{p0}=42.6$~nm and the DNA radius $a=0.85$~nm. \textbf{(b)} $(1:2)$ MgCl$_2$ salt in blue ($L_{p0}=41.4$~nm and $a=0.57$~nm) and $(2:2)$ MgSO$_4$ salt in red ($L_{p0}=39$~nm and $a=0.65$~nm). \textbf{(c)} $(2:1)$ Na$_2$SO$_3$ salt in blue ($L_{p0}=33.8$~nm $a=1.1$~nm) and $(2:1)$ Na$_2$SO$_4$ salt in red  ($L_{p0}=37.7$~nm $a=1.2$~nm). In~(a) [respectively (c)], the 3 (resp. 4) first data points are not included in the fit. Bootstraping standard errors $\Delta L_p$ are smaller than symbol size~(see SM).}
\label{f2}
\end{center}
\end{figure}

Based on these expressions for  $\xi_{\rm eff}(\kappa a,\xi)$, we compared the theoretical dsDNA persistence length to those measured by TPM. Before fitting the case of divalent ions, we have checked again the $(1:1)$ case.
Using the same experimental protocol as described in Ref.~\cite{PRL2019} (see the Appendix), we fitted a new set of data for the $(1:1)$ case using NaCl with more points at low salt concentration, i.e starting at $I=5\times 10^{-4}$~mol/L, as shown in \fig{f2}(a).
We used a home-made fitting procedure using a simulated annealing algorithm.
We thus check that the theory agrees well with the experiment, except for the first three points at low concentration, below $I_{\rm res}\simeq 10^{-3}$~mol/L. The experimental persistence length remains constant below $I_{\rm res}$, which is probably due to the fact, that, despite the rinsing of the sample, some ions stay in the solution at this residual ionic strength. The fitting parameters are $L_{p0}=42.6$~nm and $a=0.85$~nm, very similar to the ones found in Ref.~\cite{PRL2019} for another set of data.
The case of divalent counterions, using a salt of magnesium-chloride MgCl$_2$ 
is shown in \fig{f2}(b) (blue curve). Here again the theory agrees very well with the experiments on the whole ionic strength range, with fitting parameters equal to $L_{p0}=41.4$~nm and $a=0.57$~nm. The $(2:2)$ case with also a magnesium counterion (magnesium sulfate, MgSO$_4$) 
is shown in the same figure (red curve), where $L_p(I)$ is very similar to MgCl$_2$. The agreement with \eq{Lel} is extremely good with fitting parameters $L_{p0}=38.2$~nm and $a=0.64$~nm. In both cases, we obtain a fitted DNA radius lower than the structural one of 1~nm.
Finally in \fig{f2}(c) is plotted the persistence length with divalent co-ions, $(2:1)$ case, for two different salts: sodium sulfite Na$_2$SO$_3$ 
and sodium sulfate Na$_2$SO$_4$. 
The fit is very good except at low ionic strengths ($I<I_{\rm res}\simeq 1.5\times10^{-3}$~mol/L), as in the $(1:1)$ case of \fig{f2}(a). The fitting parameters are respectively $a=1.1$~nm and  $L_{p0}=33.8$~nm for Na$_2$SO$_3$ and $a=1.2$~nm and $L_{p0}=37.7$~nm for Na$_2$SO$_4$.


The first fitting parameter is the persistence length in the high salt limit, $L_{p0}$, which varies between 33.8 and 42.6~nm in the different experiments. We do not have any simple explanation for this little discrepancy. We notice however in the data some slight fluctuations at high salt which leads us to evaluate the experimental error bar on $L_{p0}$ around 3~nm. 
For the second fitting parameter, the dsDNA radius $a$, which is known to be between $0.95$~\cite{Shabarova} and $1.02$~nm~\cite{Mandelkern}, we obtain two series of data: between 0.85 and 1.2~nm respectively for the $(1:1)$ and $(1:2)$ salts, i.e. for monovalent counterions, and $\simeq 0.6$~nm for the divalent magnesium counterion [$(2:1)$ and $(2:2)$ salts]. This lower value for the dsDNA radius can be explained by the adsorption of dehydrated $\mathrm{Mg}^{2+}$ ions which are known to enter partly inside the double-helix (sequence-specific binding) as first measured by Murk Rose \textit{et al.} using NMR relaxation~\cite{Rose1980}, and then by X-ray diffraction~\cite{Subirana2003} which shows that minor and major groove binding involves H-bond interactions~\cite{Chiu2000}. Subsequent molecular dynamics simulations show the existence of binding pockets for hydrated magnesium ions with longer ion residence times~\cite{Pan2014}. This is consistent with the introduction of a lower effective dsDNA radius in the theory.

Note that in Ref.~\cite{PRL2019}, the data for divalent magnesium, calcium and putrescine counterions where fitted using the following formula $L_p = L_{p,0} + 0.238 a^{0.728} I^{-0.636}$ considering that $\xi_{\rm eff} = 0.423 \xi (\kappa R_{\rm DNA})^{0.364}$ (and neglecting the term $4\kappa a$ in \eq{Lel}). This last scaling law comes from the variational approach developed by Netz and Orland~\cite{Netz2003} adapted to the case of a solid cylinder in Ref.~\cite{Macromolecules} and fitted in the range $4\leq I\leq 600$~mmol/L. The major drawback of this approach is that it does not fit satisfactorily the $(1:1)$ salt case, contrary to the theory used in this work.

In conclusion, we have fitted experimental data of the dsDNA persistence length as a function of the concentration of $(1:1),(2:1),(1:2)$ and $(2:2)$ salts using a unique theoretical approach~\cite{TS}, and for a wide range of ionic strengths of biophysical and biotechnological interest. This approach yields a simple formula for the electrostatic persistence length, \eq{Lel}, where the central parameter is the effective linear charge density $\xi_{\rm eff}(\kappa a, \xi)$, the expression of which is obtained by matching exact asymptotic solutions of the non-linear Poisson-Boltzmann equation, in the low ($\kappa a\ll 1$) and high salt ($\kappa a\gg 1$) limits. The theoretical fits agree very well with the experimental data on an unprecedented range of ionic strength $1.5\times 10^{-3}<I<0.5$~mol/L. The dsDNA radius obtained for the monovalent counterion case is 1~nm, consistent with the actual value, whereas we obtain a lower 0.6~nm value for the $\mathrm{Mg}^{2+}$ ion. We interpret this lower value as a consequence of the adsorption of $\mathrm{Mg}^{2+}$ in the grooves of the double-helix. Understanding the impact of various ions on the DNA bending is important for the genome compaction in the cell where many divalent ions are present.

\section*{Appendix}
\label{appendix}
The DNA sample was produced by polymerase chain reaction amplification (Taq Plus from VWR) using oligos (from Sigma-Aldrich),  Biot-R1095 5'-CGGGCCTCTTCGCTATTAC-3' and Dig-F104 5'- CCGGATCAAGAGCTACCAAC-3' on pUC19 plasmid and purified with phenol chloroform~\cite{PRL2019}. 
On glass slides that constitute the top of the fluidic chamber, four holes are drilled with a sand blasting machine. Top and bottom glass slide are cleaned by immersion into a 4\% hellmanex solution at $60^\circ$C for 4~min, then sonicated in deionised water during 5 min, rinsed with a large amount of deionised water and finally dried under nitrogen flow. The cleaned slides are epoxydized by an incubation (1 h 30 min at room temperature) in isopropanol with 3-glycidoxypropyldimethoxymethylsilane 2.5\% v/v, deionised water 0.5\% v/v and N,N-dimethylbenzylamine 0.05\% v/v (Sigma-Aldrich). After treatment, the slides are sonicated in isopropanol during 5~min, and finally rinsed with a large amount of deionised water and dried under nitrogen flow. 
Square array of isolated spots of neutravidin (Invitrogen), $\sim800$~nm size separated by $\sim3~\mu$m, are formed on the bottom glass slide by micro-contact printing. For that, silicon (Si) masters with inverted patterns (AMO GmbH, Aachen, Germany) are silanized with octadecyltrichlorosilane (OTS, 98\%, Sigma-Aldrich) before their first use. PDMS (Sylgard 184, DowCorning, USA), obtained by vigorously mixing prepolymer/curing agent in a 10/1 ratio and degassing 2~min at $1300 g$ in a centrifuger, is poured on the Si masters and cured at $65^\circ$C for 8~h. PDMS stamps are cut in small pieces and inked with $20~\mu$L of a neutravidin solution (0.02~mg/ml in PBS, $p{\rm H}=7.4$) for 1~min, rinsed with deionised water and dried under nitrogen flow. The inked PDMS stamps are put in conformal contact with epoxydized coverslips for 1~min. This bottom micro-contact printed glass slide is assembled with the top glass slide drilled with four holes and a silicone spacer that forms two channels of $\sim20~\mu$L.
The internal surface of the chamber is passivated during 10~min with a PBS buffer (Euromedex) supplemented with pluronic F127 1~mg/mL and BSA 0.1~mg/mL (Sigma-Aldrich).  A 1:1 mix of 50~pM DNA and 300~nm-sized polystyrene particles (Merck) coated with anti-digoxygenin (Roche) is incubated during 30 min at $37^\circ$C and injected in the channels for an overnight incubation.
The channel under study was extensively rinsed ($\sim100$ chamber volumes) with the zero-salt-buffer composed of 1~mM HEPES set at $p{\rm H}=7.3$ by addition of NaOH, pluronic F127 1~mg/mL and BSA 0.1~mg/mL (Sigma),  then with X-salt-buffer composed of the zero-salt-buffer supplemented with various concentrations of X ions. The salts employed are NaCl (Normapur VWR), sodium sulfate (13464 Sigma), sodium thiosulfate (217263 Sigma), MgCl$_2$ (M1028 Sigma). After addition of 1~mL X-salt-buffer, the channel was left to incubate for 4~min. After flusing again $200~\mu$L  X-salt-buffer, the acquisition was performed. Subsequently, the channel was again extensively rinsed ($\sim100$ chamber volumes) with the zero-salt-buffer before any new salt condition was applied. Channel under examination underwent an increase in concentration of a single type of ion only. Experiments were repeated on different days to ensure the reproducibility of our results. Data acquisitions and analysis were carried out exactly as described in~\cite{Macromolecules}. 

The number of tracked particles lies between 325 and 2000 for each condition (salt type and ionic strength), except for NaCl experiments where it dwells between 260 and 1000. The statistical standard errors on the mean on $L_p$ (bootstrap estimate, see Ref.~\cite{Macromolecules}) are smaller than 0.9 nm (1.7 nm for NaCl).

\end{document}